\pdfoutput=1

\documentclass[11pt,twoside,a4paper,cmspaper,final,collab]{cms-tdr}
\def\svnVersion{227780}\def\cmsCernNo{2013-189}\def\cmsCernDate{\today}\def\cmsMessage{Published in Physics Letters B as \href{http://dx.doi.org/10.1016/j.physletb.2014.01.042}{\doi{10.1016/j.physletb.2014.01.042.}}}

\begin{document}\cmsNoteHeader{HIN-12-002}

\hyphenation{had-ron-i-za-tion}
\hyphenation{cal-or-i-me-ter}
\hyphenation{de-vices}

\RCS$Revision: 222081 $
\RCS$HeadURL: svn+ssh://svn.cern.ch/reps/tdr2/papers/HIN-12-002/trunk/HIN-12-002.tex $
\RCS$Id: HIN-12-002.tex 222081 2013-12-30 14:14:50Z alverson $
\newlength\cmsFigWidth
\ifthenelse{\boolean{cms@external}}{\setlength\cmsFigWidth{0.95\columnwidth}}{\setlength\cmsFigWidth{0.55\textwidth}}
\ifthenelse{\boolean{cms@external}}{\providecommand{\cmsLeft}{top}}{\providecommand{\cmsLeft}{left}}
\ifthenelse{\boolean{cms@external}}{\providecommand{\cmsRight}{bottom}}{\providecommand{\cmsRight}{right}}
\providecommand{\pp}{\Pp\Pp\xspace}
\providecommand{\ppbar}{\Pp\Pap\xspace}
\providecommand{\PbPb}{\ensuremath{\cmsSymbolFace{PbPb}}\xspace}
\providecommand{\HYDJET}{\textsc{hydjet}\xspace}

\cmsNoteHeader{HIN-12-002} % This is over-written in the CMS environment: useful as preprint no. for export versions
\title{Modification of jet shapes in PbPb collisions at $\sqrt{s_{_{\mathrm{NN}}}} = 2.76\TeV$}

\date{\today}

\abstract{
The first measurement of jet shapes, defined as the fractional transverse momentum radial distribution, for inclusive jets produced in heavy-ion collisions is presented. Data samples of \PbPb and \pp collisions, corresponding to integrated luminosities of
150\mubinv and 5.3\pbinv respectively, were collected at a nucleon-nucleon centre-of-mass energy of $\sqrt{s_{_{\mathrm{NN}}}} = 2.76\TeV$ with the CMS detector at the LHC. The jets are reconstructed with the anti-\kt
algorithm with a distance parameter $R=0.3$, and the jet shapes are measured for charged particles with transverse
momentum $\pt> 1\GeVc$. The
jet shapes measured in \PbPb collisions in different collision centralities are compared to reference distributions based on the \pp data. A centrality-dependent modification of the jet shapes is observed in the more central \PbPb collisions, indicating a redistribution of the energy inside the jet cone. This measurement provides information about the parton shower mechanism in the hot and dense medium produced in heavy-ion collisions.
}

\hypersetup{%
pdfauthor={CMS Collaboration},%
pdftitle={Modification of jet shapes in PbPb collisions at sqrt(s[NN]) = 2.76 TeV},%
pdfsubject={CMS},%
pdfkeywords={CMS, heavy ion physics, jet shapes}}

\maketitle %maketitle comes after all the front information has been supplied

 \section{Introduction}

High transverse momentum partons produced in heavy-ion collisions are
expected to lose energy while traversing the hot and dense medium
created in these collisions~\cite{Bjorken:1982tu,Gyulassy:1990ye}.
This phenomenon, known  as ``jet quenching'', was discovered at the Relativistic Heavy Ion Collider (RHIC)~\cite{Adcox:2001jp,Adler:2002xw,Adler:2003ii}
through the observation of a suppressed production of high transverse momentum (\pt) particles, and a modification of back-to-back dihadron yields~\cite{Adler:2002tq,Adams:2003im}.
These findings contributed to the conclusion that a strongly coupled quark-gluon plasma (sQGP) is being produced in nucleus-nucleus
collisions at RHIC~\cite{PHENIX,STAR,PHOBOS,BRAHMS}. The characterization of the properties of the sQGP and their evolution
with centre-of-mass energy is a central goal of the heavy-ion experimental programs at RHIC and at the
Large Hadron Collider (LHC). Despite the extensive set of RHIC measurements (see \eg~\cite{d'Enterria:2009am} for a review)
utilising high \pt particles that are presumed to be the leading fragments of jets, the jet-medium interactions
are not yet completely understood and theoretical descriptions yield significant qualitative differences in the
extracted medium properties~\cite{Bass:2008rv,Renk:2011aa,Armesto:2011ht}.

Measurements involving a full set of jet observables provide more stringent
constraints~\cite{Majumder:2010qh,Mehtar-Tani:2013pia}.
With the order of magnitude increase in the centre-of-mass energy compared to RHIC, jets with energies well above the background from the
underlying event are observed at the LHC. This has allowed the study of jet quenching
with high \pt particles~\cite{CMS:2012aa,Aamodt:2010jd}, inclusive jets~\cite{:2012is, Chatrchyan:2012gt} and jet coincidence
measurements~\cite{CMS_dijet2010,CMS_dijet,Chatrchyan:2012gt,Aad:2010bu}.

Jet quenching studies have been performed as a function of the collision centrality, defined as a fraction of the total inelastic nucleus-nucleus cross section,
with 0\% denoting the most central collisions (impact parameter $b =$ 0) and 100\% denoting the
most peripheral collisions. A strong increase in the fraction of jet pairs with largely unbalanced transverse momentum has been observed in central \PbPb collisions compared to peripheral collisions and to unquenched Monte-Carlo models~\cite{CMS_dijet2010,Aad:2010bu}.
The missing jet energy was found to be carried by soft particles that are well separated in direction from the axis of the back-to-back jets~\cite{CMS_dijet2010}.
The fragmentation patterns of the jet
constituents were found to be consistent with the fragmentation in \pp~collisions at the same nucleon-nucleon
centre-of-mass energy~\cite{Chatrchyan:2012gw} when only high \pt hadrons ($>$4\GeVc) are considered. To further elucidate the multi-gluon emission process and the in-medium shower
development~\cite{MehtarTani:2010ma,Idilbi:2008vm,CasalderreySolana:2011rz,CasalderreySolana:2011rq,Neufeld:2011yh,Blaizot:2012fh,Fickinger:2013xwa}, we study the inclusive
jet transverse-momentum profiles (shapes) in \PbPb collisions of different centralities including
low-\pt particles ($>$1\GeVc), and perform a direct comparison with the jet shapes measured in \pp collisions. The jet shapes describe how the jet transverse momentum is distributed as a function of the radial distance from the jet axis and
are expected to contain important information on the energy loss mechanism due to the medium interactions~\cite{Salgado:2003rv,Vitev,
Renk:2009hv}.
These studies complement
the previous measurements~\cite{CMS_dijet2010} in which jet-track correlations were studied in several bins of dijet asymmetry.

Jet shape measurements are challenging due to the difficulties in discriminating hadrons
originating in the parton shower from those produced by the thermally equilibrated medium or through soft processes in the underlying event. Furthermore, the lost energy may result
in softer hadrons that appear in the tail of the jet energy distribution and require large statistics for any
modification of the jet shapes to be measured reliably.
During the 2011 heavy-ion data-taking at the LHC, the Compact Muon Solenoid (CMS) experiment recorded \PbPb collisions at a nucleon-nucleon centre-of-mass energy of $\sqrt{s_{_{\mathrm{NN}}}} = 2.76\TeV$.
The integrated luminosity of 150\mubinv for the collected sample is sufficient to allow the first measurement of jet shapes in heavy-ion collisions. A reference measurement in \pp collisions at the same centre-of-mass energy has been performed using the data sample from 2013. The pp integrated luminosity of 5.3\pbinv yields a kinematic reach for the jets similar to that in the \PbPb data. The modification of the jet shapes in the \PbPb collisions is studied in five classes of collision centrality: 0--10\%, 10--30\%, 30--50\%, 50--70\% and 70--100\%.

\section{Experimental setup, triggers, and event selection}
The CMS detector~\cite{CMS_Detector} features nearly hermetic calorimetric coverage and high-resolution tracking
for the reconstruction of energetic jets and charged particles.
The calorimeters consist of a lead tungstate crystal electromagnetic calorimeter (ECAL) and a brass-scintillator hadron calorimeter (HCAL) with coverage up to $\abs{\eta}=$ 3, where $\eta = -\ln[\tan(\theta/2)]$, and $\theta$ is the polar angle relative to the counterclockwise beam direction.
The quartz/steel forward hadron calorimeters (HF) extend the calorimetry coverage in the pseudorapidity region
of   $3< \abs{\eta}< 5.2$ and are used to determine the centrality of the \PbPb collision~\cite{Chatrchyan:2011pb}.
The calorimeter cells are grouped in projective towers of granularity
$\Delta\eta\times\Delta\phi = 0.087 \times0.087$ (where $\phi$ is the azimuthal angle in radians) for the central pseudorapidities used in the jet measurement, and
have coarser segmentation (about twice as large) at forward pseudorapidity. The central calorimeters are embedded in a
superconducting solenoid with 3.8\unit{T} central magnetic field. The jet shape measurements are performed using charged particles
that are reconstructed by the CMS tracking system, located inside the calorimeter and the superconducting coil. It consists of
silicon pixel and strip layers covering the pseudorapidity range $\abs{\eta}< $2.5 and provides track reconstruction with momentum resolution of about 1-2\% up to \pt = 100\GeVc. A set of scintillator tiles, the beam scintillator counters (BSC),
are mounted on the inner side of the HF calorimeters and are used for triggering and beam-halo rejection. The BSCs cover the range
$3.23 < \abs{\eta}<4.65$.

For online event selection, CMS uses a two-level trigger system:
a level-1 (L1) and a high-level trigger (HLT). The events selected for this
analysis in \PbPb collisions are using an inclusive single-jet trigger  which
requires an L1 jet with $\pt > 52\GeVc$ and an HLT jet with $\pt >
80\GeVc$, where the jets are reconstructed using the energy deposits in the calorimeters.
The jet \pt value is corrected for the \pt-dependent
calorimeter energy response using the online implementation of the
jet-finding algorithm~\cite{JETJINST}. For \pp collisions, events are selected if they
pass an L1 jet trigger threshold of $\pt >36\GeVc$ and an HLT jet threshold of $\pt > 80\GeVc$.
The jet trigger efficiencies are evaluated using
minimum-bias-triggered events that are selected by requiring
coincidence signals registered either in the BSCs, or in the HF
calorimeters. For jets with $\pt > 100\GeVc$, both the \pp and \PbPb jet triggers are found to be fully efficient with respect
to the offline jet reconstruction, which uses a different jet algorithm based on the full detector information, as discussed in Sec.~\ref{sec:jetrecon}.

Offline selections are applied to ensure a pure sample of inelastic
hadronic collision events both in the jet-triggered and in minimum-bias-triggered events. These selections remove contamination from non-collision
beam backgrounds and from ultraperipheral collisions (UPC) that lead to an
electromagnetic breakup of one or both of the Pb nuclei. The beam-halo events are vetoed
based on the BSC timing.  To remove UPC and beam-gas events, an
offline HF coincidence of at least three towers on each side of
the interaction point is required, with a total deposited energy
of at least 3\GeV in each tower. Events with a reconstructed primary vertex based on at least two tracks
with transverse momenta above 75\MeVc are selected.
Rejection of beam-induced background events is based on the compatibility
of pixel cluster shapes with the reconstructed primary vertex. Additionally, a small number of
events with instrumental noise in the hadron calorimeter are
removed. The events are sorted according to their centrality based on the energy deposited in the HF calorimeters.
Further details on the triggering and event selections can be found in Ref.~\cite{CMS_dijet}.

\section{Jet and track reconstruction}
\label{sec:jetrecon}

Monte Carlo (MC) event generators have been used
for evaluation of the jet and track reconstruction performance, particularly
for determining the tracking efficiency, as well as the jet energy response and resolution. Jet events are generated by the
{\PYTHIA} MC generator~\cite{bib_pythia} (version 6.423, tune
Z2~\cite{Field}). These generated {\PYTHIA} events are propagated through the
CMS detector using the \GEANTfour package~\cite{GEANT4} to simulate
the detector response. In order to account for the influence of the underlying \PbPb events, the \PYTHIA events are
embedded into fully simulated \PbPb events, generated by \HYDJET~\cite{Lokhtin:2005px} (version 1.8) that is tuned to reproduce the total particle multiplicities,
charged-hadron spectra, and elliptic flow at all centralities. The embedding is done by mixing the
simulated digital signal information from \PYTHIA and \HYDJET, hereafter referred to as {\PYTHIA+\HYDJET}. These events are then propagated through the standard reconstruction and analysis chains.

For both \pp and \PbPb collisions, the analysis is based on
reconstructed jets using the anti-\kt jet-finding algorithm~\cite{bib_antikt},
with a distance parameter $R = 0.3$, utilising particle-flow (PF)
objects that combine information from all CMS detector subsystems~\cite{pas_ParticleFlow,CMS-PAS-PFT-10-002}.
The small value of $R$ is used to reduce the effects of fluctuations in the heavy-ion background. In the \PbPb data, the underlying event can affect jet-finding and distort the reconstructed jet energy. The background energy is subtracted using an iterative ``noise/pedestal'' subtraction technique as described in Ref.~\cite{Kodolova_JetReco}. The jet-finding efficiency is above
99\% for jets with $\pt > 100\GeVc$ and $\abs{\eta}< 2$. The \pt of the reconstructed jets used in the analysis
is corrected to the particle-level jet \pt based on the {\PYTHIA} generator level information, where all
stable particles ($c\tau > 1\unit{cm}$) are included to determine the particle-level jet \pt~\cite{JETJINST}. %based on the generator level information.
The uncertainty on the absolute jet energy scale in \pp collisions is
about 3\% for jet $\pt > 50\GeVc$ with $\abs{\eta}< 3$. In \PbPb collisions, due to the influence of the underlying heavy-ion
events, this uncertainty increases to about 5\%.

Charged particles with transverse momentum  $\pt > 1\GeVc$ are used to reconstruct the jet shapes.
The track finding algorithms and track selection criteria are similar
to the ones used in previous CMS publications~\cite{CMS:2012aa,Chatrchyan:2012gw}. The tracks are reconstructed starting from a ``seed'' comprising three reconstructed signals (``hits") in the silicon pixel
detector that are compatible with a helical trajectory with
minimum \pt of 0.9\GeVc originating from a selected region ($\pm$1\unit{mm} in the transverse direction, and $\pm$ 2\unit{mm} longitudinally) around the reconstructed primary vertex.
This seed is then propagated outward through subsequent tracker layers using a
combinatorial Kalman-filter algorithm~\cite{Chatrchyan:2012ta}. The \PbPb and the \pp data are reconstructed using the same procedure.
The geometric acceptance and algorithmic efficiencies are not studied separately, but considered together as an absolute efficiency.
The track selection criteria
are optimised to ensure that the contribution from misidentified tracks and secondary particles is as low as possible in the whole kinematic range for
this analysis ($\abs{\eta}< 2.3$), while preserving a reasonable reconstruction efficiency.
For the present analysis, the tracking efficiency at $\pt =1\GeVc$ is 45\% in the most central (0--10\%) \PbPb
collisions and 53\% in \pp collisions. The efficiency increases at higher \pt and is approximately constant above $\pt = 5\GeVc$, reaching 65\% (70\%) in \PbPb (\pp) collisions. The misidentified track and secondary particle contributions are typically
below 2\%  for all the samples used in this analysis, with the exception of particles with $\pt <2\GeVc$ detected at forward pseudorapidities
and in the most central \PbPb events, where the misidentified track and secondary particle contributions reach 6\% and 3\%, respectively. For each centrality class used in the analysis, the
charged-particle yields  are corrected for efficiency, misidentified track and secondary contributions on a track-by-track basis
using a detailed ($\eta$,\pt) map determined from \PYTHIA events embedded into \HYDJET background. A similar efficiency correction has been performed for \pp data with the correction factors derived from \PYTHIA. In the
simulation, the corrected reconstructed track \pt distribution agrees
to within 3\% with the generator level inclusive charged-particle distribution at any given \pt.

\section{Analysis method and systematic uncertainties}

The differential jet shape, $\rho(r)$, describes the radial distribution of transverse momentum inside the jet cone:

\begin{equation}
\rho(r) = \frac{1}{\delta r} \frac{1}{N_\text{jet}} \sum_\text{jets} \frac{\sum\limits_{\text{tracks}~\in~[r_{\mathrm{a}}, ~r_{\mathrm{b}})}{p_\text{T}^\text{track}}}{\pt^\text{jet}}
\label{eq:rho(r)}
\end{equation}

where the jet cone is divided into six annuli with radial width $\delta r = 0.05$, and each annulus has an inner radius of $r_{\mathrm{a}}= r-\delta r/2$ and outer radius of $r_{\mathrm{b}}=r+\delta r/2$.

Here $r = \sqrt{\smash[b]{(\eta_{\text{track}} -\eta_{\text{jet}})^2 + (\phi_{\text{track}} -\phi_{\text{jet}})^2} }\le 0.3$ is the reconstructed track's radial distance from the jet axis, defined by the coordinates $\eta_{\text{jet}}$ and $\phi_{\text{jet}}$.
The transverse momenta of the reconstructed track and jet are denoted $p_{\mathrm{T}}^\text{track}$ and $\pt^\text{jet}$ respectively.
After applying tracking efficiency corrections, the transverse momentum of all charged particles with $\pt > 1\GeVc$ in each annulus is summed to obtain the fraction of the total jet \pt carried by these particles. The results are
averaged over the total number of selected jets, $N_\text{jet}$.

In heavy-ion collisions, particles from the underlying event that happen to fall inside the jet cone would modify its shape.
To compensate, this contribution is subtracted following a procedure previously employed by CMS in the measurement of the jet
fragmentation function~\cite{Chatrchyan:2012gw}. To estimate the charged-particle background, a ``background cone'' is defined by reflecting the
original jet axis about $\eta=0$, while preserving its $\phi$ coordinate (``$\eta$-reflected" method).
To avoid overlap between the signal jet region and
the background cone, jets with axes in the region $|\eta_{\text{jet}}| < 0.3$ are excluded from the analysis. Larger exclusion regions, up to $|\eta_{\text{jet}}| < 0.8$ have also been studied to investigate possible biases in this procedure due to large-angle correlations between the particles originating from
different jets in the event. The size of the exclusion region is not found to be a significant source of systematic uncertainty in the jet-shape measurement.

\begin{figure*}[hpbt]
\begin{center}
\mbox{\includegraphics[width=0.95\linewidth]
{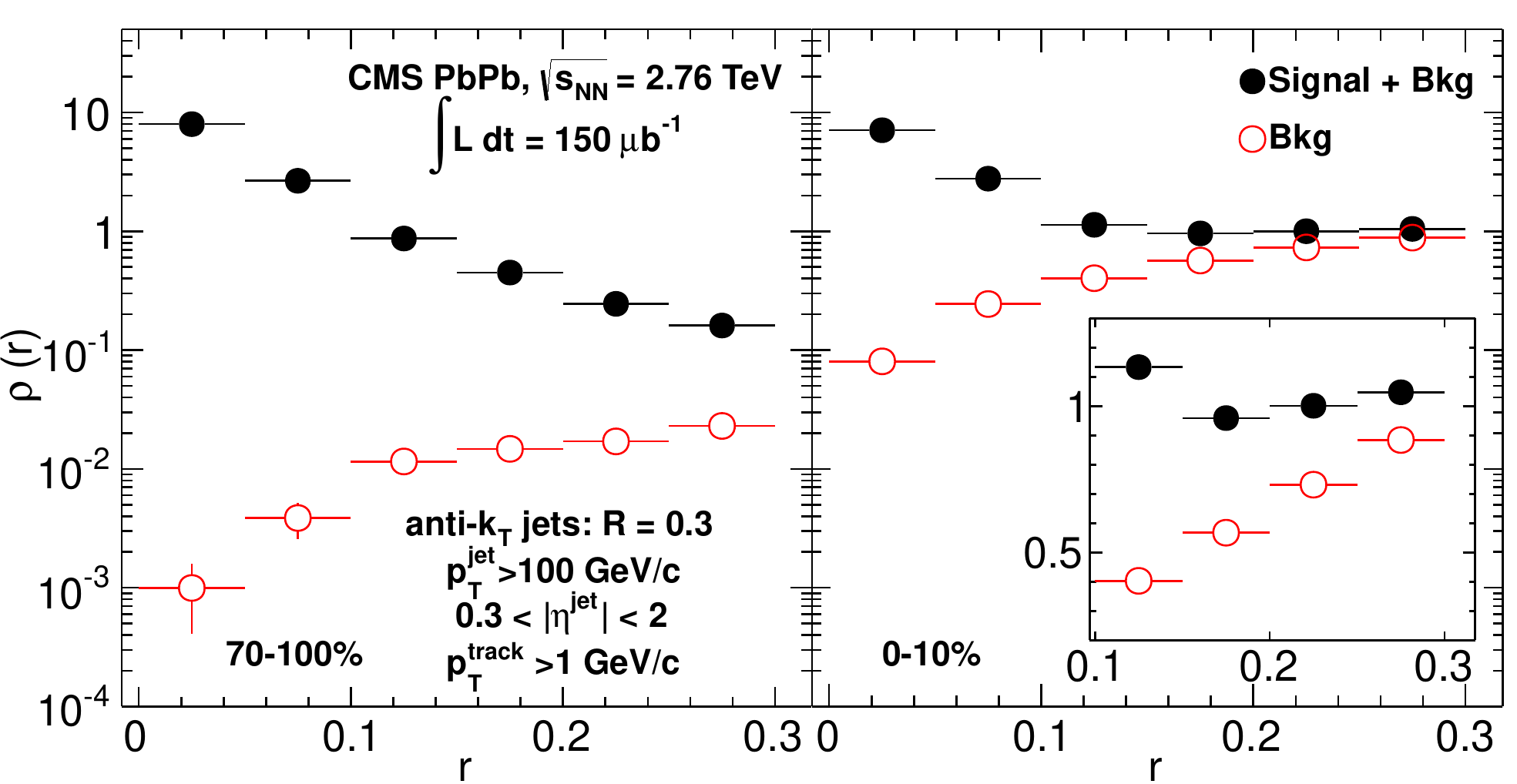}}\hfill
\caption{\label{fig:RawBkgJS}
(Color online) Differential jet shapes obtained from the jet cone before background subtraction (filled circles), and from
the ``$\eta$-reflected" background cone (open circles), as a function of the distance from the jet axis in two \PbPb centrality
intervals: 70--100\% (left) and 0--10\% (right).  The measurements use inclusive jets with $\pt^\text{jet} > 100\GeVc$ and $0.3 < \abs{\eta}< 2$, and charged particles with $\pt^\text{track} >1\GeVc$. }
\end{center}
\end{figure*}

The charged particles that are found
in the background cone are used to evaluate the background jet shape using Eq.~\ref{eq:rho(r)}, which is
then subtracted from the reconstructed jet shape that contains both signal and background particles.
After background
subtraction, the integral of $\rho(r)$ over the range $0\leq r\leq R$ is normalized to unity. The normalization factor accounts
for the average fraction of the total jet \pt carried by charged particles with $\pt >1\GeVc$.
The differential jet shapes reconstructed using all charged particles (labeled ``Signal+Bkg'') and the corresponding background distributions (labeled ``Bkg'') are shown in Fig.~\ref{fig:RawBkgJS} for the
most peripheral (70--100\%) and the most central (0--10\%) collisions. The background is a small fraction of the result ($\leq$ 1\%) in the centre of the jet but contributes a larger fraction further away from the jet axis. In peripheral events, the fraction of background at large radii is only about 15\%, but it is significantly larger ($\approx$ 85\%) in central events.

\begin{figure}[hpbt]
\begin{center}
\includegraphics[width=\cmsFigWidth]{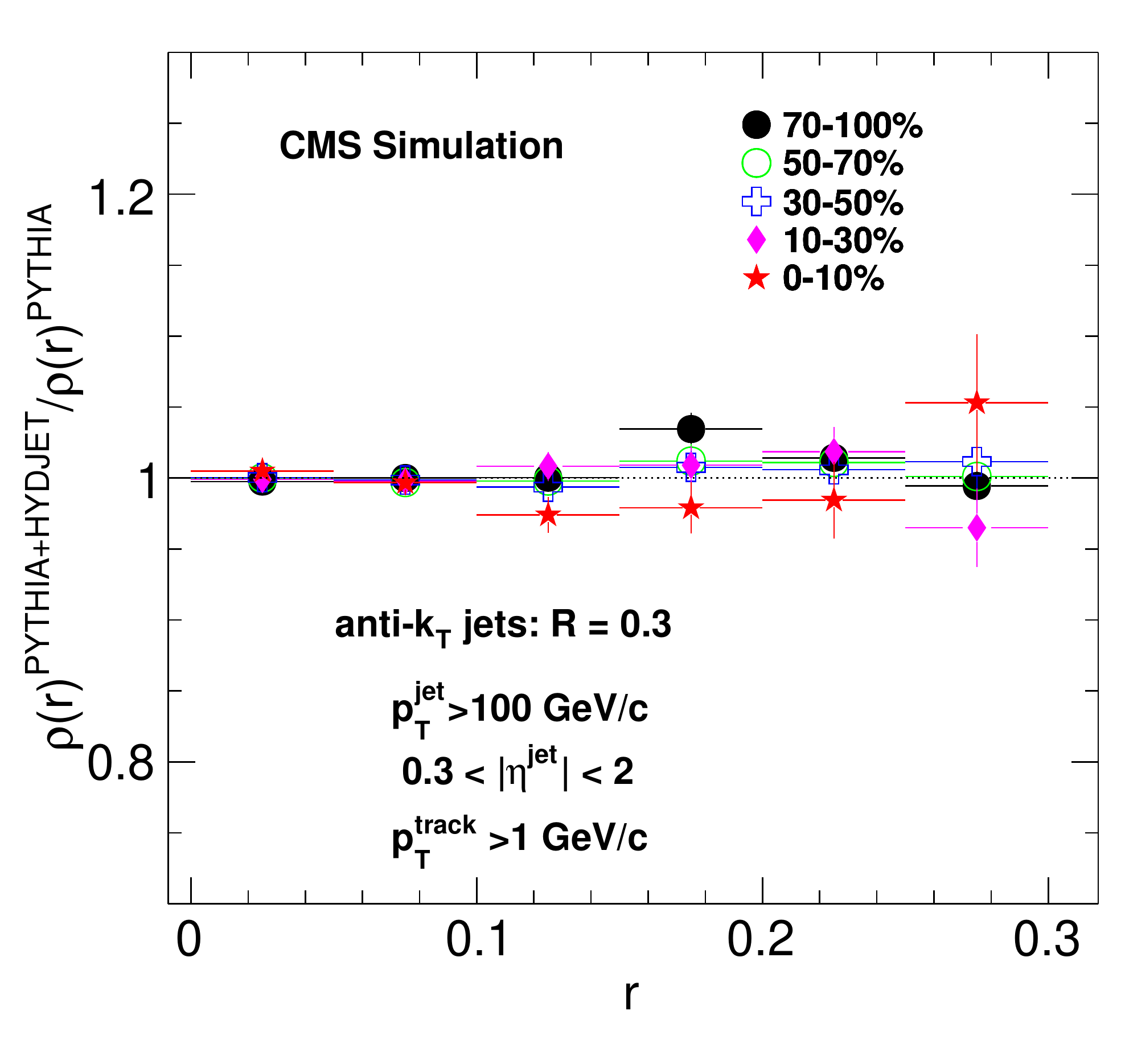}
\caption{\label{fig:MCClosure}
(Color online) Ratio of jet shapes for jets generated with \PYTHIA and embedded into heavy-ion background events simulated by
 \HYDJET, to those obtained from the \PYTHIA signal alone. The analysis uses the same background subtraction procedure
(``$\eta$-reflected") as in data. The measurements use inclusive jets with $\pt^\text{jet} > 100\GeVc$ and $0.3 < \abs{\eta}< 2$, and charged particles with $\pt^\text{track} > 1\GeVc$. }
\end{center}
\end{figure}

The  background-subtraction technique is validated using MC simulations. Jets generated with {\PYTHIA} are embedded
into heavy-ion underlying event of various centrality classes generated with the \HYDJET event generator. %events are analyzed using the same procedure as in data.
The results of the differential jet-shape measurements from embedded events are then compared to those
 obtained from a \PYTHIA jet sample at the generator level, using the same analysis procedure. The ratios of the background-subtracted shapes measured from {\PYTHIA+\HYDJET} sample and those measured in the \PYTHIA sample are shown in Fig.~\ref{fig:MCClosure} for the five centrality
classes used in the analysis. The agreement is better than 5\% even for the most central collisions, where
the background is relatively large and its fluctuations become important.

An alternative ``event-mixing'' technique is used as a cross-check of the background subtraction procedure. Minimum-bias-triggered events are
considered to be representative of the background.
Each jet-triggered event is randomly matched to ten minimum bias \PbPb events that are required to have similar global characteristics: the primary vertices have longitudinal positions within 5\unit{cm}, the relative difference in centrality is less than 25\%, and the event plane angle, as determined from the azimuthal anisotropy of the energy deposited in the HF calorimeters~\cite{Chatrchyan:2012ta}, is the same within 250\unit{mrad}.
Alternate values for these selection requirements are also studied and lead to negligible differences in the analysis results. The particles from these selected minimum bias events are used to evaluate the background jet shapes corresponding to the jets in the signal sample. The differential jet shapes obtained using this alternative background estimation are then compared to those obtained using the $\eta$-reflection technique and the ratio of the two results is used to estimate a systematic uncertainty as listed in Table~\ref{tab:sys}.

The goal of the present measurement is to study the modifications of the jet structure due to the presence of a hot
and dense medium produced in \PbPb collisions. To evaluate these modifications in each centrality interval of the \PbPb
measurements, a reference differential jet shape distribution is constructed from the \pp data, taking into account the
differences in the jet energy scale, jet momentum resolution, and the jet \pt distributions in the two systems. These
adjustments are needed to ensure that the comparison is free of detector effects and that the kinematic range of the jets included
in the comparison is the same. Based on MC studies, the reconstructed \pt of every jet in the \pp data has been shifted to remove any residual differences in the jet energy
scale between the two systems due to background fluctuations. These shifts are of the order of 1--2\%, depending on the centrality. Next, the jet \pt is smeared using a
Gaussian distribution with a standard deviation determined by the quadratic difference of the jet energy resolution
in \PbPb and \pp~collisions.
The smearing factors are derived from MC studies, which show that the jet momentum response
has little or no deviation from a Gaussian shape in both collision systems. The resulting jet \pt spectrum is
compared to the spectra in \PbPb collisions in each centrality class in order to determine a jet \pt-dependent weight. This is
applied on a jet-by-jet basis to ensure that the spectra of the jets that are included in the jet shape measurement are identical in the two
systems. This is important,  since the jet shapes depend on jet \pt.
Several parametrizations of the \pt dependence of the resolution and the jet energy scale are used to evaluate the uncertainties in this procedure.

Several sources of systematic uncertainties are considered in the measurement of the jet shapes in \PbPb collisions
and of their nuclear modification factors, $\rho(r)^{\PbPb}/\rho(r)^{\pp}$. These sources include the tracking efficiency, the background subtraction method, the jet energy resolution and the jet energy scale.  They are evaluated as a function of the
jet-track distance $r$ and are summarized in Table~\ref{tab:sys}. Uncertainties are shown for representative radius bins only, they vary smoothly in other radius bins that are not shown. Except for those due to the background subtraction, the systematic uncertainties have little or no centrality dependence. The table
shows the maximum values that typically
correspond to the most central collisions (0--10\%). The uncertainties of the \pt-dependent tracking efficiency and misidentified tracks corrections result in uncertainties in the jet shapes, especially at a large distance from the core of the jet ($r\ge 0.2$), where most of the particles have low \pt. These uncertainties are less than 7\% in the jet shape profile distribution and 4\% in the jet shape ratios, and are independent of the centrality of the collisions. The uncertainties resulting from the background subtraction are
evaluated using the relative difference in the jet shape distribution obtained with the two background subtraction procedures
described above and from the simulation studies at the generator level. In more central events (0--30\%), the background subtraction uncertainty is found to be of the order of 2\% at the core of the jet and reaches 10\% at large distances from the jet axis, where the background is more significant compared to the signal level. These uncertainties are smaller for the more peripheral events (2\% for 50--100\% and 5\% for 30--50\% centrality intervals). The uncertainties in the jet energy scale and resolution also have an impact on the measured shapes, as
jets migrating in and out of the selected jet \pt range may have a different shape~\cite{Chatrchyan:2012mec}. In \PbPb collisions,
the uncertainty from this source is of the order of 7\% at a large distance from the jet axis, and independent of the centrality of the collisions. In the jet shape ratios, these uncertainties are smaller and have been evaluated by varying the smearing and shifting
parameters used in constructing the \pp reference spectrum taking into account the relative uncertainty in the jet energy resolution
and scale in the two systems.
The systematic uncertainties from different sources are added in quadrature.

\begin{table}[htbp]
\begin{center}
\topcaption{Systematic uncertainties in the measurement of the differential jet shapes in \PbPb collisions,
and in the ratio to the \pp based reference measurement $\rho(r)^{\PbPb}/\rho(r)^{\pp}$. The
uncertainties have a small centrality dependence and the table shows the maximum values that typically
correspond to the most central collisions.}
\label{tab:sys}
\begin{tabular}{c|cccc}
\hline
\multirow{3}{*} {Source} & \multicolumn{4}{c}{ Distance from jet axis (radius) } \\ \cline{2-5}
& \multicolumn{2}{c|}{$ r \leq 0.1$} & \multicolumn{2}{c}{$r \geq 0.2$} \\ \cline{2-5}
& $\rho(r)$ & Ratio & $\rho(r)$ & Ratio \\
\hline
Background subtr.                         & 2\% & 2\% & 10\% & 10\% \\
Tracking eff.                     & 7\% & 4\% & 7\% & 4\% \\
Jet energy scale \& res.         & 2\% & 2\% & 7\% & 5\% \\
\hline
Total & 8\% & 5\% & 14\% & 12\% \\

\hline
\end{tabular}
\end{center}
\end{table}

\section{Results and discussion}

\begin{figure*}[htbp]
\begin{center}
\includegraphics[width=0.95\textwidth]
{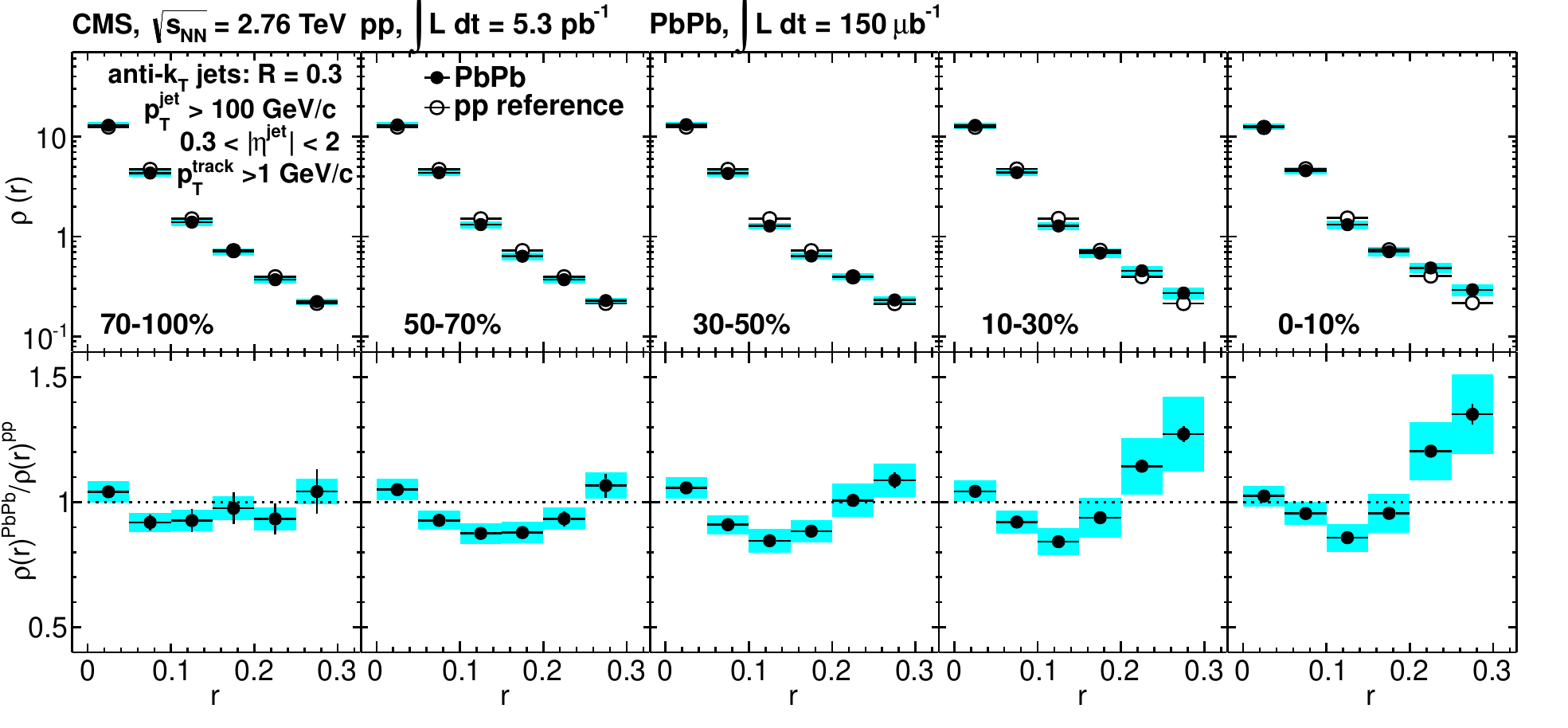}
\caption{\label{fig:JSRatio}
(Color online) Top row: Differential jet shapes in \PbPb collisions (filled circles)
 as a function of distance from the jet axis for inclusive jets with $\pt^\text{jet} >100\GeVc$ and $0.3 < \abs{\eta} < 2$ in
five \PbPb centrality intervals. The measurements use charged particles with  $\pt^\text{track} > 1\GeVc$.
The \pp-based reference shapes (with centrality-based adjustments as described in the text) are shown with open symbols.
Each spectrum is normalised to an integral of unity.
The shaded regions represent the systematic uncertainties for the measurement performed in \PbPb collisions, with the
statistical uncertainties too small to be visible.
Bottom row: Jet shape nuclear modification factors, $\rho(r)^\PbPb/\rho(r)^{\pp}$.
The error bars show the statistical uncertainties, and the shaded boxes indicate the systematic uncertainties. }
\end{center}
\end{figure*}

The differential jet shapes measured in \PbPb collisions and the reference constructed based on measurements from \pp data
are presented in the top row of Fig.~\ref{fig:JSRatio}
for five centrality classes, varying from most peripheral 70--100\% (left) to most
central 0--10\% (right). In both collision systems, 85\% of the transverse momentum is concentrated in the core
of the jet at radii $r<$ 0.1, and the small amount ($\approx$5\%) of \pt contained at radii $r> 0.2$ is carried by
low-\pt  particles. The largest differences between the jet shapes measured in the \PbPb and \pp~systems are observed
at large radii in the most central \PbPb collisions.
To quantify these modifications,
the ratios $\rho(r)^\PbPb/\rho(r)^{\pp}$ are plotted in the bottom row of Fig.~\ref{fig:JSRatio}.
Deviations from unity indicate modification of jet structure in the nuclear medium. Recall that the integrals of the
jet shapes are normalised to unity and, as a result, an excess at one distance $r$ from
the jet axis has to be compensated by a depletion at another location.
In the peripheral collisions (70--100\%), the ratio is close to unity within the uncertainties in the whole measured range,
which indicates that the radial distribution of the summed transverse momentum of the particles inside the jets is similar in the two systems. In more central \PbPb collisions (0--70\%), a
depletion is observed in the region $0.1 < r < 0.2$ with a typical value of the ratio $\rho(r)^\PbPb/\rho(r)^{\pp}$ around 0.84 and a total uncertainty of less than 7\%.
In the most central \PbPb collisions  (10--30\% and 0--10\%), an excess of transverse momentum fraction emitted at large radius $r > 0.2$ emerges,
indicating a moderate broadening of the jets in the medium.  At the largest radius $0.25 < r < 0.3$, the value of the ratio $\rho(r)^\PbPb/\rho(r)^{\pp}$ is $1.04 \pm 0.09\stat \pm 0.05\syst$ for the most peripheral collisions (70-100\%), while in the central collisions (10--30\% and 0--10\%) it increases to $1.27 \pm 0.03\stat \pm 0.15\syst$ and $1.35 \pm 0.05\stat \pm 0.16\syst$, respectively.  These observations are consistent with previous
studies in CMS which find that the energy   that the jets lose in the  medium is redistributed at large distances
from the jet axis outside the jet cone~\cite{CMS_dijet2010}. The differential study of the jet structure presented here provides
important additional information and shows that nuclear modifications are also present inside the jet cone. Qualitatively, a similar trend is predicted by theory~\cite{Vitev,Renk:2009hv} based on parton level calculations for \PbPb collisions at a different centre-of-mass energy.
It is expected that a detailed  theory-experiment comparison will be performed in the future, in which the theoretical calculations would include all experimental cuts that would influence the observed correlations, and model the effects due to the hadronization process. This comparison will contribute to our understanding of the medium properties.

\section{Summary}

The first measurement of jet shapes in \PbPb collisions at $\sqrt{s_{_{\mathrm{NN}}}} =2.76\TeV$ has been performed. The results have been compared to reference shapes measured in \pp~collisions at the same centre-of-mass energy. Inclusive jets with $\pt^\text{jet} > 100\GeVc$ and $0.3 < \abs{\eta}< 2$ have been reconstructed
using the anti-\kt algorithm with a distance parameter $R = 0.3$, and the jet shapes have been studied using charged
particles with $\pt >1\GeVc$ as a function of collision centrality.
In peripheral collisions, the shapes in \PbPb are similar to those in the \pp reference distributions. A centrality dependent modification
of the jet shapes emerges in the more central \PbPb collisions. A redistribution of the jet energy inside the cone is found,
specifically,
a depletion of jet transverse momentum fraction at intermediate radii,  $0.1 < r < 0.2$, and an excess
at large radii, $r >0.2$. These results are important for characterizing the shower evolution
in the presence of a hot and dense nuclear medium.

\section*{Acknowledgments}

We congratulate our colleagues in the CERN accelerator departments for the excellent performance of the LHC and thank the technical and administrative staffs at CERN and at other CMS institutes for their contributions to the success of the CMS effort. In addition, we gratefully acknowledge the computing centres and personnel of the Worldwide LHC Computing Grid for delivering so effectively the computing infrastructure essential to our analyses. Finally, we acknowledge the enduring support for the construction and operation of the LHC and the CMS detector provided by the following funding agencies: BMWF and FWF (Austria); FNRS and FWO (Belgium); CNPq, CAPES, FAPERJ, and FAPESP (Brazil); MES (Bulgaria); CERN; CAS, MoST, and NSFC (China); COLCIENCIAS (Colombia); MSES (Croatia); RPF (Cyprus); MoER, SF0690030s09 and ERDF (Estonia); Academy of Finland, MEC, and HIP (Finland); CEA and CNRS/IN2P3 (France); BMBF, DFG, and HGF (Germany); GSRT (Greece); OTKA and NKTH (Hungary); DAE and DST (India); IPM (Iran); SFI (Ireland); INFN (Italy); NRF and WCU (Republic of Korea); LAS (Lithuania); CINVESTAV, CONACYT, SEP, and UASLP-FAI (Mexico); MBIE (New Zealand); PAEC (Pakistan); MSHE and NSC (Poland); FCT (Portugal); JINR (Dubna); MON, RosAtom, RAS and RFBR (Russia); MESTD (Serbia); SEIDI and CPAN (Spain); Swiss Funding Agencies (Switzerland); NSC (Taipei); ThEPCenter, IPST, STAR and NSTDA (Thailand); TUBITAK and TAEK (Turkey); NASU (Ukraine); STFC (United Kingdom); DOE and NSF (USA).

Individuals have received support from the Marie-Curie programme and the European Research Council and EPLANET (European Union); the Leventis Foundation; the A. P. Sloan Foundation; the Alexander von Humboldt Foundation; the Belgian Federal Science Policy Office; the Fonds pour la Formation \`a la Recherche dans l'Industrie et dans l'Agriculture (FRIA-Belgium); the Agentschap voor Innovatie door Wetenschap en Technologie (IWT-Belgium); the Ministry of Education, Youth and Sports (MEYS) of Czech Republic; the Council of Science and Industrial Research, India; the Compagnia di San Paolo (Torino); the HOMING PLUS programme of Foundation for Polish Science, cofinanced by EU, Regional Development Fund; and the Thalis and Aristeia programmes cofinanced by EU-ESF and the Greek NSRF.

\bibliography{auto_generated}   % will be created by the tdr script.

\cleardoublepage \appendix\section{The CMS Collaboration \label{app:collab}}\begin{sloppypar}\hyphenpenalty=5000\widowpenalty=500\clubpenalty=5000\textbf{Yerevan Physics Institute,  Yerevan,  Armenia}\\*[0pt]
S.~Chatrchyan, V.~Khachatryan, A.M.~Sirunyan, A.~Tumasyan
\vskip\cmsinstskip
\textbf{Institut f\"{u}r Hochenergiephysik der OeAW,  Wien,  Austria}\\*[0pt]
W.~Adam, T.~Bergauer, M.~Dragicevic, J.~Er\"{o}, C.~Fabjan\cmsAuthorMark{1}, M.~Friedl, R.~Fr\"{u}hwirth\cmsAuthorMark{1}, V.M.~Ghete, N.~H\"{o}rmann, J.~Hrubec, M.~Jeitler\cmsAuthorMark{1}, W.~Kiesenhofer, V.~Kn\"{u}nz, M.~Krammer\cmsAuthorMark{1}, I.~Kr\"{a}tschmer, D.~Liko, I.~Mikulec, D.~Rabady\cmsAuthorMark{2}, B.~Rahbaran, C.~Rohringer, H.~Rohringer, R.~Sch\"{o}fbeck, J.~Strauss, A.~Taurok, W.~Treberer-Treberspurg, W.~Waltenberger, C.-E.~Wulz\cmsAuthorMark{1}
% ...
\vskip\cmsinstskip
% ...
1:~~Also at Vienna University of Technology, Vienna, Austria\\
2:~~Also at CERN, European Organization for Nuclear Research, Geneva, Switzerland\\
% ...

\end{sloppypar}
\end{document}